\documentclass[12pt,a4paper]{article}
\usepackage[latin2]{inputenc}
\usepackage{t1enc}
\usepackage{amscd}
\usepackage{amsmath}
\usepackage{amssymb}
\usepackage{amsxtra}
\usepackage{setspace}
\usepackage{tabularx}
\usepackage{graphics}
\usepackage{fancyhdr}
\usepackage[thmmarks]{ntheorem-hyper}

\theoremheaderfont{\bf}\theorembodyfont{\it}\theoremsymbol{\fbox{}}\theoremstyle{plain}\newtheorem{Thm}{Theorem}
\theoremheaderfont{\bf}\theorembodyfont{\sl}\theoremsymbol{\setlength{\unitlength}{1mm}\begin{picture}(2.5,2.5)\linethickness{2.5mm}\put(0,1.25){\line(1,0){2.5}}\end{picture}}\theoremstyle{plain}\newtheorem{Prf}{Proof}
\theoremheaderfont{\bf}\theorembodyfont{\it}\theoremsymbol{}\theoremstyle{plain}\newtheorem{Def}[Thm]{Definition}
\theoremheaderfont{\bf}\theorembodyfont{\sl}\theoremsymbol{}\theoremstyle{plain}\newtheorem{Rem}[Thm]{Remark}
\theoremheaderfont{\bf}\theorembodyfont{\it}\theoremsymbol{\fbox{}}\theoremstyle{plain}\newtheorem{Lem}[Thm]{Lemma}
\theoremheaderfont{\bf}\theorembodyfont{\sl}\theoremsymbol{}\theoremstyle{plain}
\theoremheaderfont{\bf}\theorembodyfont{\it}\theoremsymbol{\fbox{}}\theoremstyle{plain}\newtheorem{Col}[Thm]{Corollary}

\newcommand{\Tr}{\mathop{\mathrm{Tr}}\nolimits}

\begin{document}

\pagestyle{empty}

\begin{center}

\vspace*{5mm}

{\begin{spacing}{1}\LARGE\textbf{Mathematical Clarification of General Relativistic Variational Principles}\end{spacing}}

\vspace*{5mm}

{\large András LÁSZLÓ}

{Eötvös University, Budapest, Hungary}

{\texttt{laszloa@szofi.elte.hu}}

\vspace*{5mm}


\begin{abstract}
\noindent
In this paper a mathematically precise global (i.e. not the usual local) 
approach is presented to the variational principles 
of general relativistic classical field theories.

Problems of the classic (usual) approaches are also discussed in comparison.

The aim of developing a global approach is to provide a possible tool for 
future efforts on proving global existence theorems of field theoretical solutions.
\end{abstract}

\end{center}

\pagestyle{fancy}

\setcounter{page}{1}

\lhead{}\chead{\scriptsize{Mathematical Clarification of General Relativistic Variational Principles}}\rhead{}

\lfoot{}\cfoot{\thepage}\rfoot{}

\section{Introduction}

As one can find out from physics and mathematics literature, the known 
variational formulations of 
general relativistic classical field theories can be divided into three classes. 
These classes 
differ in the definition of the action functional (in the definition of the 
integration domain of the action functional), and in the notion of 
variation.\footnote{The concept of variation of the action functional is 
a notion of a kind of derivative. Some of 
the approaches use one-parameter families of field configurations to define 
the variation of the action functional. This derivative-like notion resembles 
to the Gateaux derivative (directional derivative). Other approaches 
use more adequate notion for the variation, which corresponds to the Fréchet 
derivative (classic derivative of a map between a normed affine space and 
a topological vector space, based on the notion of $\mathrm{ordo}$ functions).}

The first class of approaches specify the action functional as an integral 
of a Lagrange form\footnote{We call the product of the Lagrangian density and 
the volume form the Lagrange form.} over the spacetime manifold. This approach 
is mathematically ill defined, as the action would diverge for some 
quite physical (e.g. stationary) field configurations, in general. (This fact 
can be shown explicitly on specific examples.)

The second class specifies the action as the integral of the Lagrange form on a 
given compact subset of the spacetime manifold. The variation is then defined 
by using one-parameter families of field configurations (see e.g. [5], [8]). 
This means a Gateaux-like notion of derivative. 

The third class defines the action functional as the integral of the Lagrange form 
on the spacetime domain between two given time-slice. As time-slices may be 
noncompact in general, certain fall-off properties have to be introduced on the 
field configurations in order to make sense of the action (otherwise, the action 
could diverge for specific configurations). After specifying appropriate 
regularity conditions, one can define a natural $C^k$-type 
supremum norm equivalence class (for some nonnegative integer $k$), and the 
variation is simply defined as the (Fréchet) derivative of the 
action functional (with respect to this norm equivalence class).

All the three approaches have certain undesired properties. 
The first formulation, as pointed out, is mathematically ill defined. 
The second one is a local kind of 
formulation, and the notion of variation is Gateaux-like. The Gateaux-like 
derivative is a much weaker notion than the Fréchet derivative: the most powerful 
tools in differential calculus (e.g. the Taylor formulas, some critical point theorems 
etc.) only work with Fréchet type derivative.\footnote{It is also a well known 
fact, that the Fréchet differentiability property of a function at a point is a 
much stronger condition than the Gateaux (directional) differentiability. 
Although, in finite dimensions, the continuous Gateaux 
differentiability on an open set is known to be 
equivalent to the continuous Fréchet differentiability, this is not true in 
infinite dimensions (i.e. in our case).} 
Apart from this argument of mathematical inelegance, we will discuss a further 
problem of this approach in \emph{Section 3.2.2} (the argument of non-constructiveness), 
which is in connection with boundary value problems. 
The third approach is semi-global, as it 
is global in spatial sense, but local in the timelike direction, furthermore the 
notion of variation is Fréchet type, which is potentially more powerful, when 
trying to prove e.g. some critical point theorems. However, there are several difficulties 
concerning this kind of formulation, which are discussed in the end of \emph{Subsection 3.2.2} 
(the problem of spacetime splitting and the problem of spatial fall-off conditions).

Inspired by the above mentioned problems, we developed a global kind of 
approach, which uses Fréchet type notion of variation. The domain of integration 
can be viewed as the conformal compactification of the arising spacetime model, 
and the notion of variation is simply the (Fréchet) derivative of the action 
functional, with respect to the natural $C^k$-type 
supremum norm equivalence class (for some nonnegative integer $k$) on the field 
configurations.

The presented approach resembles to Palatini formulation of general 
relativity, as the covariant derivation is taken to be an 
independent dynamical field quantity. The formalism can also handle theories 
using covariant derivations with nonvanishing torsion. A main result concerning 
theories with nonvanishing torsion can be found in \emph{Section 4}, in 
\emph{Theorem 15} 
(which presents the field equations and boundary constraints for the case of nonzero torsion).

The class of spacetime manifolds, 
which can be generated by this formulation, are the 
orientable, asymptotically simple models\footnote{In the definition of asymptotic 
simpleness, we do not include the condition of asymptotic emptiness.}.

A global approach, which uses Fréchet type notion of variation, can possibly be 
applied in future as a tool of proving global existence theorems in general 
relativistic field theories.

\section{Building up a field theory by using variational principles}

\subsection{Base manifold}

Let us take a real $C^3$ manifold $M$, which is orientable. 
$M$ will be called the \emph{base manifold}.\footnote{The 
base manifold is going to be the manifold, 
where the integration is carried out in order to define the action functional. In 
the classic formalism, 
the base manifold plays the role of spacetime manifold. However, in \emph{Section 4} and \emph{5}, 
we shall carry out an argument that the base manifold should not be 
directly interpreted as the spacetime manifold, but as the so called conformal 
compactification of the arising spacetime model. Thus, we do not refer to the 
base manifold as spacetime manifold.} 
Let us use the following notation: $m:=\dim{M}$. 
In the followings, we will denote by $\mathbb{R}$ the real numbers, 
and by $\mathbb{N}$ the positive integers.

\subsection{Field quantities}

As usually in a classical field theory, the field quantities will be sections of a 
fixed real vector bundle over the base manifold, and covariant derivations 
over the vector bundle.\footnote{This kind of notion of field 
quantities is used in the Palatini type formulation of general relativistic 
variational problems. In our approach, the torsion of the covariant derivation 
is not assumed to be zero \emph{a priori}.}

If $W(N)$ is a real $C^{\breve{k}}$ ($\breve{k} \in \{0,\dots,k\}$) vector bundle over a real $C^k$ manifold $N$, 
then the $C^l$ ($l\in \{0,\dots,\breve{k}\}$) sections of it will be denoted by ${\Gamma}^{l}(W(N))$. 
Furthermore let $D^{l}(W(N))$ be the space of $C^l$ covariant derivations 
over $W(N)$. Let $X(N)$ be a real $C^{\bar{k}}$ ($\bar{k} \in \{0,\dots,k\}$) vector bundle over $N$. 
We use the natural injection of $D^{\breve{l}}(W(N)){\times}D^{\bar{l}}(X(N))$ into 
$D^{l}(W(N){\times}X(N))$ ($\breve{l},\bar{l} \in \{0,\dots,\mathrm{min}(\breve{k},\bar{k})\}$ and 
$l \in \{0,\dots,\mathrm{min}(\breve{l},\bar{l})\}$).

\begin{Rem}
Let $F(N)$ be $N{\times}\mathbb{R}$ as a real $C^k$ vector bundle. 
Let $T(N)$ be the tangent vector bundle of $N$ as a real $C^k$ vector bundle. 
Let $W(N)$ be a real $C^{\breve{k}}$ ($\breve{k} \in \{0,\dots,k\}$) vector bundle. 
A $C^l$ ($l\in \{0,\dots,\breve{k}\}$) covariant derivation in $D^{l}(T(N)){\times}D^{l}(W(N))$ 
can be uniquely extended to a $C^l$ covariant derivation over 
all the mixed tensor and cross products of $F(N)$, $T(N)$, $W(N)$ and their duals by 
requiring the Leibniz rule, the commutativity with contraction, and the correspondence 
to the exterior derivation on $F(N)$. We can refer to this unique extension 
by the original covariant derivation, because they determine each other 
uniquely. For example if $s\in\Gamma^{2}(W(N))$ and $\nabla\in D^{2}(T(M)){\times}D^{2}(W(N))$, 
then $\nabla(\nabla(s))$ is well defined, if $k$ and $\breve{k}$ is greater or equal to $2$.

The $C^l$ covariant derivations in $D^{l}(T(N)){\times}D^{l}(W(N))$ form an affine 
space over the vector space of 
${\Gamma}^{l-1}\bigl(T^{\ast}(N){\otimes}\bigl((T(N){\otimes}T^{\ast}(N)){\times}(W(N){\otimes}W^{\ast}(N))\bigr)\bigr)$ sections, 
that is over the vector space of the so called $C^{l-1}$ \emph{diagonal Christoffel tensor fields} 
of $T(N){\times}W(N)$. This means, that if we subtract two such covariant derivation, 
their action corresponds to the action of a $C^{l-1}$ Christoffel tensor field 
of $T(N)$ (that is, to a $C^{l-1}$ section of $T^{\ast}(N){\otimes}T(N){\otimes}T^{\ast}(N)$) 
on the sections of $T(N)$, and to the action of a $C^{l-1}$ Christoffel 
tensor field of $W(N)$ (that is, to a $C^{l-1}$ section of $T^{\ast}(N){\otimes}W(N){\otimes}W^{\ast}(N)$) 
on the sections of $W(N)$. 
This fact follows from the basic properties of the covariant derivations: 
we refer to textbooks e.g. [5] and [8].
\end{Rem}

Let us fix a $C^3$ real vector bundle $V(M)$ over $M$. 
Let us introduce a sub fiber bundle $\breve{V}(M)$ of the vector bundle $V(M)$, 
with the same fiber dimension as $V(M)$ (thus, for each $p\in M$ the fiber 
$\breve{V}_{p}(M)$ is a sub manifold of $V_{p}(M)$ with dimension $\mathrm{dim}(V_{p}(M))$). 
Let $\breve{D}^{3}(T(M),V(M))$ be a closed sub affine space 
of the affine space $D^{3}(T(M)){\times}D^{3}(V(M))$, where the topology is understood 
to be the topology defined in \emph{Definition 8} in \emph{Subsection 3.1}. 
The \emph{field variables} of 
the theory are going to be the elements of ${\Gamma}^{3}(\breve{V}(M)){\times}\breve{D}^{3}(T(M),V(M))$, 
that is the covariant derivations are also dynamical.

\subsection{The Lagrange form}

We introduce a central notion of the variational principles: the Lagrange form. 
It is going to replace the notion of Lagrangian density function of the classic 
formalism. This notion is well known in literature, but there is no generally 
accepted label for it. (In the classic formalism, the Lagrange form can be obtained 
as the product of the Lagrangian density and the volume form.)

Let us take a map 
\[\begin{array}{c}\mathbf{dL}:\Gamma^3\Bigl(\breve{V}(M)\Bigr){\times}\Gamma^2\Bigl(T^{\ast}(M){\otimes}V(M)\Bigr){\times}\Gamma^2\Bigl(\overset{2}{\wedge}T^{\ast}(M){\otimes}\bigl((T^{\ast}(M){\otimes}T(M)){\times}(V^{\ast}(M){\otimes}V(M))\bigr)\Bigr)\\
\rightarrow \Gamma^{2}\Bigl(\overset{m}{\wedge}T^{\ast}(M)\Bigr)\end{array}\]
which is \emph{pointwise}, that is 
\[\begin{array}{c}{\forall}p{\in}M:\\
{\forall}v,v^{'}{\in}\Gamma^3(\breve{V}(M)),w,w^{'}{\in}\Gamma^2(T^{\ast}(M){\otimes}V(M)),\\
r,r^{'}{\in}\Gamma^2\Bigl(\overset{2}{\wedge}T^{\ast}(M){\otimes}\bigl((T^{\ast}(M){\otimes}T(M)){\times}(V^{\ast}(M){\otimes}V(M))\bigr)\Bigr):\\
(v(p)=v^{'}(p) \text{ and } w(p)=w^{'}(p) \text{ and } r(p)=r^{'}(p)) \Longrightarrow \mathbf{dL}(v,w,r)(p)=\mathbf{dL}(v^{'},w^{'},r^{'})(p).\end{array}\]
Given such a map $\mathbf{dL}$, for every $p{\in}M$ we can naturally define the map 
\[\begin{array}{c}\mathbf{dL}_{p}: \breve{V}_{p}(M){\times}\bigl(T^{\ast}_{p}(M){\otimes}V_{p}(M)\bigr){\times}\Bigl(\overset{2}{\wedge}T^{\ast}_{p}(M){\otimes}\bigl((T^{\ast}_{p}(M){\otimes}T_{p}(M)){\times}(V^{\ast}_{p}(M){\otimes}V_{p}(M))\bigr)\Bigr)\\
\rightarrow \overset{m}{\wedge}T^{\ast}_{p}(M)\end{array}\]
with the restriction of $\mathbf{dL}$ (this new function maps between 
finite dimensional vector spaces). If for every $p{\in}M$ $\mathbf{dL}_{p}$ is 
$C^2$, then we call $\mathbf{dL}$ a \emph{Lagrange form}. 
(The above requirements mean, that a Lagrange form can also be viewed as a $C^2$ 
fiber bundle homomorphism.)

\begin{Rem}
Let us take a Lagrange form $\mathbf{dL}$. Let us denote the partial derivative 
of $\mathbf{dL}_{p}$ in its $r$-th variable ($r\in\{1,2,3\}$) by $D_{r}\mathbf{dL}_{p}$ ($p{\in}M$). 
Let us take any section 
\[(v,w,r){\in}\Gamma^3(\breve{V}(M)){\times}\Gamma^2(T^{\ast}(M){\otimes}V(M)){\times}\Gamma^2\Bigl(\overset{2}{\wedge}T^{\ast}(M){\otimes}\bigl((T^{\ast}(M){\otimes}T(M)){\times}(V^{\ast}(M){\otimes}V(M))\bigr)\Bigr).\]
Then the derivative $D_{r}\mathbf{dL}_{p}(v_{p},w_{p},r_{p})$ ($r\in\{1,2,3\}$) can be viewed as 
a tensor at $p{\in}M$ of the appropriate type, because it is a linear map 
between the appropriate vector bundle fibers at $p{\in}M$. 
Furthermore, the tensor field defined by 
$p{\mapsto}D_{r}\mathbf{dL}_{p}(v_{p},w_{p},r_{p})$ ($r\in\{1,2,3\}$) is $C^1$. 
This follows from the following facts. The map $\mathbf{dL}_{p}$ is $C^2$ 
for every $p{\in}M$. Furthermore, the map $p\mapsto\mathbf{dL}_{p}(v_{p},w_{p},r_{p})$ is $C^2$ 
for every $(v,w,r)$ section as above. 
Therefore, by taking a coordinate chart on an open subset of $M$ and trivializations of the appropriate 
vector bundles over it, we see that the function $\mathbf{dL}$ (taken in coordinates) 
is $C^2$ in its manifold coordinate variable, and is $C^2$ in its vector bundle fiber coordinate 
variable. 
Thus, we have that $\mathbf{dL}$ (taken in coordinates) possesses the $C^2$ property.\footnote{This follows 
from the following theorem: given a map between a finite product of finite dimensional 
vector spaces and a finite dimensional vector space, then it is $C^1$ 
if and only if it is partially $C^1$ all in its variables.} 
Therefore, any of its partial derivatives are $C^1$: for example the partial 
derivative, which corresponds to $D_{r}\mathbf{dL}$ ($r\in\{1,2,3\}$) is also 
$C^1$ (in coordinates). By this fact, the $C^1$ property of the tensor field 
$p{\mapsto}D_{r}\mathbf{dL}_{p}(v_{p},w_{p},r_{p})$ ($r\in\{1,2,3\}$) is implied 
over an arbitrary coordinate neighborhood, thus on the whole manifold.
\end{Rem}

\subsection{The action functional}

The central notion of variational principles is the action functional. 
If one is trying to find an elegant formulation of classical field theories, 
the key step is the proper definition of the action functional. The action 
is the integral of the Lagrange form on the base manifold or on a properly 
specified subset of it.\footnote{The definitions used by classic literature 
differ here: as pointed out earlier, there are three kinds of definitions, and 
all the three approaches have certain problems. 
The alternative definition, which we give, mostly 
resembles to the second approach, as we define the action functional (in the 
case of a noncompact base manifold) 
as a real valued Radon measure on the Baire quasi-$\sigma$-ring of the base 
manifold (i.e. on the 
quasi-$\sigma$-ring, generated by the compact subsets of the base manifold).}

As a Lagrange form $\mathbf{dL}$ is volume form field valued, given a section 
\[(v,w,r){\in}\Gamma^3(\breve{V}(M)){\times}\Gamma^2(T^{\ast}(M){\otimes}V(M)){\times}\Gamma^2\Bigl(\overset{2}{\wedge}T^{\ast}(M){\otimes}\bigl((T^{\ast}(M){\otimes}T(M)){\times}(V^{\ast}(M){\otimes}V(M))\bigr)\Bigr),\]
we can integrate the volume form field $\mathbf{dL}(v,w,r)$ all over $M$, if it 
is integrable. 
If $M$ is compact, every continuous volume form field is integrable on $M$.

\begin{Def}
Let $M$ be a compact base manifold, and $\mathbf{dL}$ a Lagrange form. 
Then the \emph{action functional} defined by the Lagrange form is 
\[S: \Gamma^3(\breve{V}(M)){\times}\breve{D}^3(T(M),V(M)) \rightarrow \mathbb{R}, (v,\nabla) \mapsto S_{v,\nabla}:=\int\limits_{M}\mathbf{dL}(v,\nabla{v},F_{\nabla}),\]
where $F_{\nabla}$ is the curvature tensor of the covariant derivation $\nabla$.
\end{Def}

Unfortunately, if $M$ is noncompact, we cannot proceed with the straightforwardness 
as in the compact case.
If we would like to extend our formalism to a noncompact base manifold, 
we should proceed otherways. 
A possible way to realize the action functional, over a noncompact base manifold, could be 
to define it as a real valued Radon measure on the subsets of the manifold. 
If we follow this idea, we can make the following definition.

\begin{Def}
Let the base manifold $M$ be noncompact.
If \[(v,\nabla) \in \Gamma^3(\breve{V}(M)){\times}\breve{D}^3(T(M),V(M))\] and $K$ is a 
compact set in $M$, 
then let us define $S_{v,\nabla}(K):=\int\limits_{K}\mathbf{dL}(v,\nabla{v},F_{\nabla})$. 
The map $K \mapsto S_{v,\nabla}(K)$ uniquely extends to a real valued Radon measure on 
the Baire quasi-$\sigma$-ring 
of $M$. Let $\mathrm{Rad}(M,\mathbb{R})$ be the real vector space of the real valued Radon 
measures on the Baire quasi-$\sigma$-ring of $M$. Then, 
the \emph{action functional} is defined as the Radon measure valued map
\[S: \Gamma^3(\breve{V}(M)){\times}\breve{D}^3(T(M),V(M)) \rightarrow \mathrm{Rad}(M,\mathbb{R}), (v,\nabla) \mapsto S_{v,\nabla}.\]
Of course, this definition is also meaningful for the 
compact case, and the action in the compact case can be expressed as $(v,\nabla) \mapsto S_{v,\nabla}(M)$.
\end{Def}

\section{The field equations as Euler-Lagrange equations}

\subsection{Natural distribution topologies on the sections of a vector bundle}

Let us take a real $C^{\breve{k}}$ vector bundle $W(N)$ over the real $C^k$ manifold $N$ ($\breve{k}\in\{0,\dots,k\}$). 
Then we can define $C^l$ 
norm fields ($l\in\{0,\dots,\breve{k}\}$) on $\Gamma^{l}(W(N))$. Let us take a map 
\[\Vert\cdot\Vert: \Gamma^{l}(W(N)) \rightarrow \Gamma^{0}(F(N)), s \mapsto \Vert s \Vert,\]
which is pointwise, that is 
\[\forall p{\in}N: \forall s,s^{'}\in\Gamma^{l}(W(N)): s^{'}(p)=s(p) \Rightarrow \Vert s^{'}\Vert(p)=\Vert s\Vert(p)\]
holds (this means, that it can be viewed as a $C^{0}$ fiber bundle homomorphism). 
If for every $p{\in}N$ the map 
$\Vert\cdot\Vert_{p}: W_{p}(N) \rightarrow \mathbb{R}$, naturally defined by the restriction of 
$\Vert\cdot\Vert$, is a norm, then we call $\Vert\cdot\Vert$ a $C^l$ norm field. 
It is a fact that every $C^{\breve{k}}$ vector bundle over a $C^k$ manifold admits $C^l$ 
norm fields: by the paracompactness of manifolds\footnote{A manifold is paracompact as a consequence of its definition.}, there 
are $C^l$ Riemann metric tensor fields on the given vector bundle, 
and they naturally give rise to $C^l$ norm fields by taking the pointwise norms generated by them 
(but not all $C^l$ norm fields can be formulated in this way).

\begin{Lem}
Let $N$, $W(N)$ be as above. If $\Vert\cdot\Vert$ and $\Vert\cdot\Vert^{'}$ are $C^l$ norm fields 
on $W(N)$, then there exists a positive $c\in\Gamma^{k}(F(N))$, such that $\Vert\cdot\Vert^{'}\leq c\Vert\cdot\Vert$.
\end{Lem}

\begin{Prf}
The proof is based on the paracompactness of the differentiable manifolds and on the equivalence of norms 
on a finite dimensional vector space.

Let us take a locally finite atlas $((U_{i},\varphi_{i},f_{i}))_{i \in I}$ of $N$ with partition of unity, such that 
every $U_{i}$ ($i \in I$) has compact closure. Let us denote the closure of a set $U$ with $\overline{U}$. 
Let $n$ be the dimension of the fibers of $W(N)$. 
Let us fix a trivialization $(e_{i,j})_{j \in \{1,\dots,n\}}$ of $W(N)$ over each 
chart $(U_{i},\varphi_{i})$ ($i \in I$).

As a consequence of the equivalence of norms on a finite dimensional vector space, 
for every $p \in N$ there is a positive number $c_{p}$, such that $\Vert\cdot\Vert^{'}_{p} \leq c_{p} \Vert\cdot\Vert_{p}$. 
Furthermore, $c_{p}$ can be chosen to be $\underset{s_{p} \in W_{p}(N)\setminus\{0_{p}\}}{\mathrm{sup}} \frac{\Vert s_{p} \Vert^{'}_{p}}{\Vert s_{p} \Vert_{p}}$.

It is easily checked, that the equality 
\[\underset{p \in \overline{U_{i}}}{\mathrm{sup}} \; \underset{s_{p} \in W_{p}(N)\setminus\{0_{p}\}}{\mathrm{sup}} \; \frac{\Vert s_{p} \Vert^{'}_{p}}{\Vert s_{p} \Vert_{p}}= 
\underset{p \in \overline{U_{i}}}{\mathrm{sup}} \; \underset{S \in \mathbb{R}^{n}, |S|=1}{\mathrm{sup}} \; \frac{\Vert \sum\limits_{j=1}^{n}S_{j}e_{i,j} \Vert^{'}(p)}{\Vert \sum\limits_{j=1}^{n}S_{j}e_{i,j} \Vert(p)}\]
holds ($i \in I$). The rightside is a finite positive number, because it can be 
viewed as the maximum of a positive valued continuous function over the compact 
manifold $\overline{U_{i}}\times\mathbb{S}^{n-1}$ ($\mathbb{S}^{n-1}$ is the $n-1$ dimensional unit sphere). 
Let us denote this positive number by $c_{i}$. 
Then: $\Vert\cdot\Vert^{'}\leq c_{i}\Vert\cdot\Vert$ 
holds over $\overline{U_{i}}$ ($i \in I$).

As a consequence of the previous inequality: $f_{i}\Vert\cdot\Vert^{'}\leq c_{i}f_{i}\Vert\cdot\Vert$ 
holds all over the manifold $N$ for each $i \in I$, as a consequence of the fact that $f_{i}$ is nonnegative and $\mathrm{supp}(f_{i})\subset U_{i}$. 
The sum $\sum\limits_{i \in I}c_{i}f_{i}$ (which has only finite nonzero terms in a small neighborhood of every point, as a consequence of the local finiteness of the atlas) 
is a positive valued $C^{k}$ function, furthermore the sum $\sum\limits_{i \in I}f_{i}$ (which also has only finite nonzero terms in a small neighborhood of every point) 
is $1$ by definition. Therefore $\Vert\cdot\Vert^{'}\leq \left(\sum\limits_{i \in I}c_{i}f_{i}\right)\Vert\cdot\Vert$ holds, 
so the lemma is proved.
\end{Prf}

We call this property the \emph{equivalence} of $C^l$ norm fields, in the analogy 
of the equivalence of norms on a finite dimensional vector space.

Let us take a $C^{\tilde{k}}$-type covariant derivation $\nabla$ from $D^{\tilde{k}}(T(N)){\times}D^{\tilde{k}}(W(N))$ ($\tilde{k}\in \{0,\dots,\breve{k}\}$). If we take $C^{\tilde{k}-l}$ norm fields 
\[\Vert\cdot\Vert_{l}: \Gamma^{\tilde{k}-l}((\overset{l}{\otimes}T^{\ast}(N)){\otimes}W(N)) \rightarrow \Gamma^{0}(F(M))\]
for each $l\in\{0,\dots,\tilde{k}\}$, then we can formulate the quantity 
\[\sum_{l=0}^{\tilde{k}}\Vert\nabla^{(l)}\cdot\Vert_{l}: \Gamma^{\tilde{k}}(W(N)) \rightarrow \Gamma^{0}(F(M)), s \mapsto \sum_{l=0}^{\tilde{k}}\Vert\nabla^{(l)}s\Vert_{l}\]
(this is not a norm field, because it is not pointwise, but it is a similar quantity).

\begin{Lem}
If we choose norm fields $(\Vert\cdot\Vert_{l})_{l\in\{0,\dots,\tilde{k}\}}$, and two $C^{\tilde{k}}$ covariant derivations $\nabla$ and $\nabla^{'}$ as above, then 
there exists a positive $C^k$ function $c$ over $N$, such that 
$\sum\limits_{l=0}^{\tilde{k}}\Vert{\nabla^{\; '}}^{(l)}\cdot\Vert_{l}\leq c\sum\limits_{l=0}^{\tilde{k}}\Vert\nabla^{(l)}\cdot\Vert_{l}$.
\end{Lem}

\begin{Prf}
This is a consequence of the following facts:
\begin{enumerate}
\item the covariant 
derivation $\nabla^{'}$ can be expressed as the sum of $\nabla$ and a $C^{\tilde{k}-1}$ diagonal Christoffel tensor field,
\item the triangle inequality of the norms,
\item the composition of a norm with a linear map is a semi-norm,
\item the sum of a norm and a semi-norm is a norm,
\item \emph{Lemma 5}.
\end{enumerate}
\end{Prf}

\begin{Col}
If we take norm fields $(\Vert\cdot\Vert_{l})_{l\in\{0,\dots,\tilde{k}\}}$, $(\Vert\cdot\Vert^{'}_{l})_{l\in\{0,\dots,\tilde{k}\}}$ and 
covariant derivations $\nabla$ and $\nabla^{'}$ as above, then 
there exists a positive $C^k$ function $c$ over $N$, such that 
\[\sum_{l=0}^{\tilde{k}}\Vert{\nabla^{\; '}}^{(l)}\cdot\Vert^{\; '}_{l}\leq c\sum_{l=0}^{\tilde{k}}\Vert\nabla^{(l)}\cdot\Vert_{l}.\]
\end{Col}

\begin{Prf}
This is a consequence of \emph{Lemma 6} and \emph{Lemma 5}.
\end{Prf}

\emph{Corollary 7} lets us define the notions of distribution topologies on 
the vector space of sections of vector bundles.

\begin{Def}
Let ${\cal E}^{\tilde{k}}(W(N)):=\Gamma^{\tilde{k}}(W(N))$ with the natural real vector space structure. 
Let us choose a class of norm fields $(\Vert\cdot\Vert_{l})_{l\in\{0,\dots,\tilde{k}\}}$ and 
a covariant derivation $\nabla$ as before. 
Let $\psi,\varphi_{n}\in{\cal E}^{\tilde{k}}(W(N))$ ($n\in\mathbb{N}$), then the sequence 
$n\mapsto\varphi_{n}$ \emph{converges to} $\psi$ \emph{
in ${\cal E}^{\tilde{k}}$-sense}, if and only if 
the function $\sum\limits_{l=0}^{\tilde{k}}\Vert\nabla^{(l)}(\psi-\varphi_{n})\Vert_{l}$ 
converges to zero uniformly on every compact set. This notion uniquely characterizes 
a topology on ${\cal E}^{\tilde{k}}(W(N))$, which is called the \emph{
${\cal E}^{\tilde{k}}$-topology}. 
Note, that as a consequence of \emph{Corollary 7}, this notion is independent 
of the chosen norm fields and covariant derivation.
\end{Def}

\begin{Def}
Let ${\cal D}^{\tilde{k}}(W(N))$ be the set of elements of $\Gamma^{\tilde{k}}(W(N))$, which have compact support. 
${\cal D}^{\tilde{k}}(W(N))$ has a natural real vector space structure. 
Let us choose a class of norm fields $(\Vert\cdot\Vert_{l})_{l\in\{0,\dots,\tilde{k}\}}$ and 
a covariant derivation $\nabla$ as before.
Let $\psi,\varphi_{n}\in{\cal D}^{\tilde{k}}(W(N))$ ($n\in\mathbb{N}$), then the sequence 
$n\mapsto\varphi_{n}$ \emph{converges to} $\psi$ \emph{
in ${\cal D}^{\tilde{k}}$-sense}, if and only if there exists a compact set $K$, such that $\forall n\in\mathbb{N}:\mathrm{supp}(\psi-\varphi_{n})\subset K$ 
and the function $\sum\limits_{l=0}^{\tilde{k}}\Vert\nabla^{(l)}(\psi-\varphi_{n})\Vert_{l}$ 
uniformly converges to zero. This notion uniquely characterizes 
a topology on ${\cal D}^{\tilde{k}}(W(N))$, 
which is called the \emph{
${\cal D}^{\tilde{k}}$-topology}. 
Note, that as a consequence of \emph{Corollary 7}, this notion is independent 
of the chosen norm fields and covariant derivation.
\end{Def}

\begin{Def}
Let $N$ be compact. Then ${\cal E}^{\tilde{k}}(W(N))={\cal D}^{\tilde{k}}(W(N))$, and the 
${\cal E}^{\tilde{k}}$ and ${\cal D}^{\tilde{k}}$-topologies are the same. Furthermore, if 
we choose a class of norm fields $(\Vert\cdot\Vert_{l})_{l\in\{0,\dots,\tilde{k}\}}$ and 
a covariant derivation $\nabla$ as before, then the quantity 
$\underset{N}{\mathrm{sup}}\sum\limits_{l=0}^{\tilde{k}}\Vert\nabla^{(l)}\cdot\Vert_{l}$ 
is always finite, and this is a complete norm on ${\cal E}^{\tilde{k}}(W(N))$. If we take an other 
class of norm fields $(\Vert\cdot\Vert^{'}_{l})_{l\in\{0,\dots,\tilde{k}\}}$ and a 
covariant derivation $\nabla^{'}$ as before, then there is a positive 
number $c$, such that 
\[\underset{N}{\mathrm{sup}}\sum_{l=0}^{\tilde{k}}\Vert{\nabla^{\; '}}^{(l)}\cdot\Vert^{\; '}_{l}\leq c\; \underset{N}{\mathrm{sup}}\sum_{l=0}^{\tilde{k}}\Vert\nabla^{(l)}\cdot\Vert_{l},\]
that is the two norms are equivalent. Let us call them \emph{
$C^{\tilde{k}}$-norms} 
of ${\cal E}^{\tilde{k}}(W(N))$. It is easily seen, that the ${\cal E}^{\tilde{k}}$-topology is the same 
as the topology generated by any $C^{\tilde{k}}$-norm on ${\cal E}^{\tilde{k}}(W(N))$. 
These are the consequences of \emph{Corollary 7} and of the fact, that the 
continuous real valued functions over a compact manifold are bounded.
\end{Def}

\subsection{The derivative of action functional}

Let us take $(v,\nabla),(v^{'},\nabla^{'})\in\Gamma^3(\breve{V}(M)){\times}\breve{D}^3(T(M),V(M))$, 
then we can express the primed quantities as 
$v^{'}=v+{\delta}v$, $\nabla^{'}=\nabla+{\delta}C$, where the vector field ${\delta}v$ 
is $C^3$ and the diagonal Christoffel tensor field ${\delta}C$ is $C^2$.

\begin{Thm}
Let $(v,\nabla)$ and $({\delta}v,{\delta}C)$ be the above quantities, and $K$ a 
compact subset of $M$. Then 
\[S_{v+{\delta}v,\nabla+{\delta}C}(K)=S_{v,\nabla}(K)+\]
\[\int\limits_{K}\Bigl(D_{1}\mathbf{dL}(v,\nabla v,F_{\nabla}){\delta}v+D_{2}\mathbf{dL}(v,\nabla v,F_{\nabla})(\nabla{\delta}v+{\delta}Cv+{\delta}C{\delta}v)+
D_{3}\mathbf{dL}(v,\nabla v,F_{\nabla})(F_{\nabla+{\delta}C}-F_{\nabla})\Bigr)\]
\[+\int\limits_{K}\frac{1}{2}\begin{bmatrix} {\delta}v & ({\nabla}{\delta}v+{\delta}Cv+{\delta}C{\delta}v) & (F_{\nabla+{\delta}C}-F_{\nabla}) \end{bmatrix}
\cdot\begin{bmatrix}D^{(2)}\mathbf{dL}\end{bmatrix}(v+{\delta}v^{'},{\nabla}v+{\delta}A^{'},F_{\nabla}+{\delta}F^{'})\cdot\]
\[\begin{bmatrix} {\delta}v \cr ({\nabla}{\delta}v+{\delta}Cv+{\delta}C{\delta}v) \cr (F_{\nabla+{\delta}C}-F_{\nabla})\end{bmatrix}\]
for some sections ${\delta}v^{'}$, ${\delta}A^{'}$, ${\delta}F^{'}$, where for each $p\in K$ there 
exists a number $c_{p}\in ]0,1[$, such that ${\delta}v^{'}_{p}=c_{p}{\delta}v_{p}$, 
${\delta}A^{'}_{p}=c_{p}(\nabla{\delta}v|_{p}+{\delta}Cv|_{p}+{\delta}C{\delta}v|_{p})$ and ${\delta}F^{'}_{p}=c_{p}(F_{\nabla+{\delta}C}|_{p}-F_{\nabla}|_{p})$. 
Here the action of a $T^{\ast}(M){\otimes}\bigl((T(M){\otimes}T^{\ast}(M)){\times}(V(M){\otimes}V^{\ast}(M))\bigr)$ type tensor field ${\delta}C$ 
on a $V(M)$ type tensor field ${\delta}v$ is defined by the contraction of the 
projection to the $T^{\ast}(M){\otimes}V(M){\otimes}V^{\ast}(M)$ component of 
${\delta}C$ by ${\delta}v$.
\end{Thm}

\begin{Prf}
This is a simple consequence of the Taylor formula for one dimensional vector space valued $C^2$ functions, 
applied to the Lagrange form, in every point of $K$: 
\[\mathbf{dL}(v+{\delta}v,(\nabla+{\delta}C)(v+{\delta}v),F_{\nabla+{\delta}C})=\mathbf{dL}(v,{\nabla}v,F_{\nabla})+\]
\[\Bigl(D_{1}\mathbf{dL}(v,\nabla v,F_{\nabla}){\delta}v+D_{2}\mathbf{dL}(v,\nabla v,F_{\nabla})(\nabla{\delta}v+{\delta}Cv+{\delta}C{\delta}v)+
D_{3}\mathbf{dL}(v,\nabla v,F_{\nabla})(F_{\nabla+{\delta}C}-F_{\nabla})\Bigr)+\]
\[\frac{1}{2}\begin{bmatrix} {\delta}v & ({\nabla}{\delta}v+{\delta}Cv+{\delta}C{\delta}v) & (F_{\nabla+{\delta}C}-F_{\nabla}) \end{bmatrix}
\cdot\begin{bmatrix}D^{(2)}\mathbf{dL}\end{bmatrix}(v+{\delta}v^{'},{\nabla}v+{\delta}A^{'},F_{\nabla}+{\delta}F^{'})\cdot\]
\[\begin{bmatrix} {\delta}v \cr ({\nabla}{\delta}v+{\delta}Cv+{\delta}C{\delta}v) \cr (F_{\nabla+{\delta}C}-F_{\nabla}) \end{bmatrix}\]
holds, with the notations in the statement of the theorem.

We have to show only that the term under the last integral (the term with the second derivatives) is integrable. 
This follows from the simple fact: we can express it via the terms with lower derivatives by using the Taylor formula. 
The terms with lower derivatives are $C^1$ by \emph{Remark 2}. Therefore, the term with the second derivatives 
is $C^1$, although the function $p\mapsto c_{p}$ need not be even measurable. 
So, the last term is integrable on $K$.
\end{Prf}

\subsubsection{The case of compact base manifold}

Our intention is to interpret the linear term of the expression in \emph{Theorem 11} 
as a kind of derivative of the action functional. 
If $M$ is compact, this can be realized quite straightforward, by using \emph{Definition 10}.

Let us assume that $M$ is compact, without boundary. Let us take the standard topology on $\mathbb{R}$. 
Let us fix a $C^3$-norm on ${\cal E}^{3}(V(M))$ and a $C^2$-norm on 
\[{\cal E}^2\Bigl(T^{\ast}(M){\otimes}\bigl((T(M){\otimes}T^{\ast}(M)){\times}(V(M){\otimes}V^{\ast}(M))\bigr)\Bigr).\]
Then we can fix a power $q$ (which is a real number greater or equal to $1$, or infinity) and 
take the $L^q$ product norm of these norms on the product space. 
With this, the space 
\[\Gamma^3(V(M)){\times}\bigl(D^3(T(M)){\times}D^3(V(M))\bigr)\]
forms an affine space over the normed space 
\[{\cal E}^{3}(V(M))\times{\cal E}^2\Bigl(T^{\ast}(M){\otimes}\bigl((T(M){\otimes}T^{\ast}(M)){\times}(V(M){\otimes}V^{\ast}(M))\bigr)\Bigr).\]
In this sense, we can take the derivative of the action functional. The derivative is 
independent of the chosen $C^3$ and $C^2$-norms, and of the power $q$ (that is, of the way of 
forming of the product norm), 
because the notion of derivative depends only on the equivalence class of norms.\footnote{If 
we choose other $C^3$ and $C^2$-norms on the previous spaces, and take an other power $q$ for 
forming an $L^q$ product norm, then we get a new norm, which is equivalent to 
the previous one.}

\begin{Thm}
The action functional $S$ is continuously differentiable, and its derivative 
at given $(v,\nabla)$ is the continuous linear map 
$({\delta}v,{\delta}C) \mapsto$ \[\int\limits_{M}\Bigl(D_{1}\mathbf{dL}(v,\nabla v,F_{\nabla}){\delta}v+D_{2}\mathbf{dL}(v,\nabla v,F_{\nabla})(\nabla{\delta}v+{\delta}Cv)+D_{3}\mathbf{dL}(v,\nabla v,F_{\nabla})2\nabla\wedge{\delta}C\Bigr),\]
where the wedge in the expression $\nabla\wedge{\delta}C$ means antisymmetrization 
in the $T^{\ast}(M)$ variable of $\nabla$ and the first $T^{\ast}(M)$ variable of ${\delta}C$ 
in the expression $\nabla{\delta}C$.
\end{Thm}

\begin{Prf}
Note that this expression is the linear term from the expression in \emph{Theorem 11}. 
First, we have to show that this linear map is a continuous linear map, and that the 
remaining bilinear term from \emph{Theorem 11} is an $\mathrm{ordo}$ function. By the differentiability properties of 
$\mathbf{dL}$, and the compactness of $M$, these facts are direct consequences of Lebesgue theorem. 
Finally, the derivative function of $S$ is continuous, for the same reason.
\end{Prf}

\begin{Thm}
The derivative of the action functional $S$ can be expressed 
at given $(v,\nabla)$ as the continuous linear map 
\[({\delta}v,{\delta}C) \mapsto \int\limits_{M}\Bigl(D_{1}\mathbf{dL}(v,\nabla v,F_{\nabla}){\delta}v-(\nabla{\cdot}D_{2}\mathbf{dL}(v,\nabla v,F_{\nabla})){\delta}v-(\Tr{T_{\nabla}}{\cdot}D_{2}\mathbf{dL}(v,\nabla v,F_{\nabla})){\delta}v\Bigr)+\]
\[\int\limits_{M}\Bigl(D_{2}\mathbf{dL}(v,\nabla v,F_{\nabla}){\delta}Cv-2(\nabla{\cdot}\widehat{D_{3}\mathbf{dL}}(v,\nabla v,F_{\nabla})){\delta}C-2(\Tr{T_{\nabla}}{\cdot}\widehat{D_{3}\mathbf{dL}}(v,\nabla v,F_{\nabla})){\delta}C\Bigr).\]
Here $T_{\nabla}$ is the torsion tensor of $\nabla$, $\Tr{T_{\nabla}}$ denotes the contraction of the second $T^{\ast}(M)$ and the $T(M)$ variable of $T_{\nabla}$. 
The hat in the $\widehat{D_{3}\mathbf{dL}}$ expression means 
antisymmetrization in the first two $T(M)$ variables. 
Finally, $\cdot$ means contraction of the $T^{\ast}(M)$ variable of $\nabla$ and 
the first $T(M)$ variable of the tensor quantity after it, or 
the contraction of the $T^{\ast}(M)$ variable of $\Tr{T_{\nabla}}$ and 
the first $T(M)$ variable of the tensor quantity after it, respectively.

Let us call the equality of the above map and the derivative of $S$ 
the \emph{Euler-Lagrange relation}, and let us call the above map the 
\emph{Euler-Lagrange map}.
\end{Thm}

\begin{Prf}
We simply make transformations of the expression in \emph{Theorem 12}.

The term $D_{3}\mathbf{dL}(v,\nabla v,F_{\nabla})2\nabla\wedge{\delta}C$ is equal to 
$2\widehat{D_{3}\mathbf{dL}}(v,\nabla v,F_{\nabla})\nabla{\delta}C$, this is easily seen for example by using Penrose abstract indices: 
\[D_{3}^{ab}\mathbf{dL}(v,\nabla v,F_{\nabla})(\nabla_{a}{\delta}C_{b}-\nabla_{b}{\delta}C_{a})=\Bigl(D_{3}^{ab}\mathbf{dL}-D_{3}^{ba}\mathbf{dL}\Bigr)(v,\nabla v,F_{\nabla})\nabla_{a}{\delta}C_{b},\]
where the indices $a,b$ indicate the $T(M)$ or $T^{\ast}(M)$ variables in question.

By the Leibniz rule 
\[D_{2}\mathbf{dL}(v,\nabla v,F_{\nabla})\nabla{\delta}v+2\widehat{D_{3}\mathbf{dL}}(v,\nabla v,F_{\nabla})\nabla{\delta}C=\]
\[\nabla\cdot\Bigl(D_{2}(\mathbf{dL}(v,\nabla v,F_{\nabla}){\delta}v\Bigr)+\nabla\cdot\Bigl(2\widehat{D_{3}\mathbf{dL}}(v,\nabla v,F_{\nabla}){\delta}C\Bigr)-\]
\[\Bigl(\nabla\cdot D_{2}(\mathbf{dL}(v,\nabla v,F_{\nabla})\Bigr){\delta}v-\Bigl(\nabla\cdot 2\widehat{D_{3}\mathbf{dL}}(v,\nabla v,F_{\nabla})\Bigr){\delta}C\]
is true.

The sum of the first two terms on the rightside of the equation can be written in 
the form $\nabla\cdot\mathbf{dA}$, where $\mathbf{dA}\in\Gamma^{1}(T(M){\otimes}\overset{m}{\wedge}T^{\ast}(M))$ 
(that is $\mathbf{dA}$ is a $C^1$ volume form valued vector field). Let us take an 
other covariant derivation $\tilde{\nabla}$ on the tensor bundles of $T(M)$, which is the Levi-Civita 
covariant derivation of some semi-Riemannian metric tensor field $\tilde{g}$ over $M$. 
If $C\in\Gamma^{3}(T^{\ast}(M){\otimes}T(M){\otimes}T^{\ast}(M))$ 
is the Christoffel tensor field on $T(M)$ of $\nabla$ relative to $\tilde{\nabla}$, then 
one can figure out the fact, that $\nabla\cdot\mathbf{dA}=\tilde{\nabla}\cdot\mathbf{dA}+(\Tr_{1}C-Tr_{2}C)\cdot\mathbf{dA}$, 
where $\Tr_{1}C$ denotes the contraction of the first $T^{\ast}(M)$ and the $T(M)$ variable of 
$C$, and $\Tr_{2}C$ denotes the contraction of the second $T^{\ast}(M)$ and the $T(M)$ variable of 
$C$. It is easier to follow the previous statement in Penrose abstract indices: 
$\nabla_{a}\mathbf{dA}^{a}=\tilde{\nabla}_{a}\mathbf{dA}^{a}+(C_{ba}^{b}-C_{ab}^{b})\mathbf{dA}^{a}$, 
because $\nabla_{a}t^{b}=\tilde{\nabla}_{a}t^{b}+C_{ac}^{b}t^{c}$ 
is valid for a tangent vector field $t$, and $\nabla_{a}\mathbf{dv}=\tilde{\nabla}_{a}\mathbf{dv}-C_{ab}^{b}\mathbf{dv}$ 
is true for a volume form field $\mathbf{dv}$. As $\tilde{\nabla}$ is a Levi-Civita covariant 
derivation, the $(\Tr_{1}C-Tr_{2}C)$ 
quantity corresponds to $-\Tr{T_{\nabla}}$, because the torsion of $\tilde{\nabla}$ vanishes by definition. 
Thus, we can write $\nabla\cdot\mathbf{dA}=\tilde{\nabla}\cdot\mathbf{dA}-\Tr{T_{\nabla}}\cdot\mathbf{dA}$.

The term with the torsion corresponds to the term with the torsion in the statement of the theorem. 
To prove the theorem, we only have to show, that the integral of $\tilde{\nabla}\cdot\mathbf{dA}$ 
is zero.

Let us use the fact that $M$ is orientable: there exists a nowhere zero $C^3$ volume form field $\mathbf{dv}$. 
As the vector space of the volume forms, at a given point, is one dimensional, then 
we can uniquely define a nowhere zero section $\mathbf{dp}$ of the dual volume form bundle, such that 
at every point, $\mathbf{dp}$ maps $\mathbf{dv}$ into $1$. By using coordinate charts, 
it can be seen, that $\mathbf{dp}$ is also $C^3$. Then by contracting the quantity 
$\mathbf{dp}{\otimes}\mathbf{dA}$ in the volume form and dual volume form variables, 
one can define a $C^1$ vector field (as $\mathbf{dA}$ is $C^1$). Let us denote this by $\mathbf{dA/dv}$. With 
the introduced notation, one has $\mathbf{dA}=\mathbf{dv}{\otimes}(\mathbf{dA/dv})$.

Let $\tilde{\mathbf{dv}}$ be one of the two 
canonical volume form fields associated to $\tilde{g}$. Then $\tilde{\nabla}\tilde{\mathbf{dv}}=0$ holds, 
which implies by the Leibniz rule: $\tilde{\nabla}\cdot\mathbf{dA}=\tilde{\mathbf{dv}}\, \tilde{\nabla}\cdot(\mathbf{dA/\tilde{dv}})$. 
It is a theorem, that if $X$ is a tangent vector field, then $\tilde{\mathbf{dv}}\, \tilde{\nabla}{\cdot}X=m\, \mathrm{d}(X.\tilde{\mathbf{dv}})$, where 
$m$ is the dimension of $M$, 
$\mathrm{d}$ means the exterior derivation, and the dot $.$ means contraction of $X$ 
with the first $T^{\ast}(M)$ variable of $\tilde{\mathbf{dv}}$ (see for example [5]). 
By using the observations in the previous paragraph, we can state that 
$(\mathbf{dA/\tilde{dv}}).\mathbf{\tilde{dv}}=\Tr{\mathbf{dA}}$, where $\Tr$ 
means the contraction of the $T(M)$ variable of $\mathbf{dA}$ with the 
first $T^{\ast}(M)$ variable of $\mathbf{dA}$. We get, that the expression $\tilde{\mathbf{dv}}\, \tilde{\nabla}\cdot(\mathbf{dA/\tilde{dv}})$ 
is independent of the choice of the semi-Riemannian metric tensor field $\tilde{g}$, and it is 
equal to $\mathrm{d}(m\Tr{\mathbf{dA}})$, the exterior derivative of the $(m-1)$-form field 
$m\Tr{\mathbf{dA}}$.
The integral of this term vanishes as a consequence of Gauss theorem, because $M$ is a compact manifold 
without boundary. 
So the formula, stated in the theorem, is valid.
\end{Prf}

\begin{Rem}
If $M$ is a compact manifold with boundary, then the presented statements 
remain true, but the Euler-Lagrange map in \emph{Theorem 13} has an extra term, 
which is a boundary integral, as a consequence of the Gauss theorem. 
Namely, the derivative of $S$ at given $(v,\nabla)$ is the continuous linear map 
\[({\delta}v,{\delta}C) \mapsto \int\limits_{M}\Bigl(D_{1}\mathbf{dL}(v,\nabla v,F_{\nabla}){\delta}v-(\nabla{\cdot}D_{2}\mathbf{dL}(v,\nabla v,F_{\nabla})){\delta}v-(\Tr{T_{\nabla}}{\cdot}D_{2}\mathbf{dL}(v,\nabla v,F_{\nabla})){\delta}v\Bigr)+\]
\[\int\limits_{M}\Bigl(D_{2}\mathbf{dL}(v,\nabla v,F_{\nabla}){\delta}Cv-2(\nabla{\cdot}\widehat{D_{3}\mathbf{dL}}(v,\nabla v,F_{\nabla})){\delta}C-2(\Tr{T_{\nabla}}{\cdot}\widehat{D_{3}\mathbf{dL}}(v,\nabla v,F_{\nabla})){\delta}C\Bigr)+\]
\[m\int\limits_{\partial{M}}\Tr\Bigl(D_{2}(\mathbf{dL}(v,\nabla v,F_{\nabla}){\delta}v+2\widehat{D_{3}\mathbf{dL}}(v,\nabla v,F_{\nabla}){\delta}C\Bigr),\]
where $\partial{M}$ is the boundary of $M$.
\end{Rem}

As a summary, we can define a classical field theory over a compact base manifold 
$M$ (with or without boundary) as a quartet $(M,V(M),\mathbf{dL},S)$, where 
$V(M)$ is a vector bundle as in the text, $\mathbf{dL}$ is a Lagrange form, and 
$S$ is the action functional, defined by $\mathbf{dL}$. The field equation 
is the equation 
\[\Bigl((v,\nabla)\in\Gamma^3(\breve{V}(M)){\times}\breve{D}^3(T(M),V(M))\Bigr)\ ?\quad DS(v,\nabla)=0,\]
where $DS$ denotes the derivative of $S$. Let us call $DS$ the \emph{Euler-Lagrange functional}.

\subsubsection{The case of noncompact base manifold}

If the base manifold is noncompact, the vector spaces of sections of a vector bundle do not 
have natural normed space structure, they only have natural ${\cal E}$ or ${\cal D}$ 
distribution topologies.

Let us take a map $Q:R\rightarrow S$, where $R$ and $S$ 
are some spaces. To be able to define the classical (Fréchet) notion of derivative of $Q$, 
the space $R$ has to be a normed affine space, and $S$ has to be a topological 
vector space (in order to be able to define the notion of $\mathrm{ordo}$ functions).

In the case of the action functional, when the base manifold is 
noncompact, the first space is the space $\Gamma^3(V(M)){\times}\bigl(D^3(T(M)){\times}D^3(V(M))\bigr)$, 
which forms an affine space over the topological vector space 
${\cal E}^{3}(V(M))\times{\cal E}^2\Bigl(T^{\ast}(M){\otimes}\bigl((T(M){\otimes}T^{\ast}(M)){\times}(V(M){\otimes}V^{\ast}(M))\bigr)\Bigr)$. 
The second space is the space of real valued Radon measures $\mathrm{Rad}(M,\mathbb{R})$, 
which is a vector space, and it possesses a natural topology, uniquely characterized 
by the following notion of limes: a sequence of 
Radon measures converges to a Radon measure, if both evaluated on any fixed compact set of $M$, 
the sequence of values (real numbers) converges to the value of the given 
Radon measure (real number). (This is the \emph{pointwise}, or the \emph{setwise} topology 
on the Radon measures.)

As we see, when $M$ is noncompact, the notion of the derivative of $S$ cannot 
be defined: the obstruction is that $\Gamma^3(V(M)){\times}\bigl(D^3(T(M)){\times}D^3V(M))\bigr)$ 
only has a natural \emph{topological} affine space structure, instead of a natural 
\emph{normed} affine space structure. Thus, if we want to proceed in the noncompact case, 
and want to define a similar quantity 
to an Euler-Lagrange functional, we can not interpret it as a (Fréchet) derivative.

There are known constructions, which are based on a formulation popular in physics literature, even in mathematical physics literature (see e.g. [5], [8]). 
It defines $(v,\nabla)\mapsto DS_{v,\nabla}(K)$ by using one-parameter families 
of field configurations, which are fixed on the boundary of a fixed compact set $K$ 
with smooth boundary. (We will refer to these formulations as \emph{one-parameter family 
formulations}.) To define the field equations, they take a covering 
$(K_{i})_{i\in I}$ of $M$ with such compact sets, and on every set they require $DS(K_{i})=0$ ($i\in I$). 
It can be proved, that $DS(K_{i})=0$ means Euler-Lagrange equations over the interior of the 
given compact set $K_{i}$ ($i\in I$), so after all, over the whole spacetime manifold $M$. 
This statement is true, but we have one more 
constraint: the field values are fixed on the system of boundaries $(\partial{K_{i}})_{i\in I}$.

As one can see, the one-parameter family construction is quite cumbersome. 
Furthermore, it is not constructive in the following sense. Let us fix a compact set 
with sooth boundary, and a field configuration on the boundary. If the Euler-Lagrange 
equations are first order hyperbolic (e.g. when the Dirac equation is part of the field 
equations), then generally there is 
no such field configuration on the compact set, which satisfies the field 
equations and has the (arbitrarily) chosen boundary values. Thus, one can not generate 
solutions inside a compact set by specifying (arbitrary) boundary values.

There are constructions known in physics literature, which are defined otherways. 
We shall refer to these as \emph{time-slice constructions}. 
These assume a cylindric base manifold, i.e. 
a manifold diffeomorphic to $\mathbb{R}\times C$ for some manifold $C$ (which will be referred 
as space or time-slice). The action is defined as the integral of the Lagrange form 
on the domain between two specific time-slice. 
Certain spatial fall-off properties have to be introduced in order to be able 
to define the action functional, if $C$ is not compact. The Euler-Lagrange functional is then defined 
as the derivative of the action with respect to appropriate $C^k$ supremum norms (for some $k\in \mathbb{N}$), 
similar to the case of compact base manifold. The problem is: how to formalize 
the spatial regularity conditions. In the literature this problem is carefully 
overlooked, if possible. The most self-suggesting solution seems to be to 
introduce a global coordinate system on $C$ (this is, of course, not always possible), 
and treat the fall-off properties with respect 
to the coordinates. This method would be quite inelegant (as it refers 
to global coordinate chart), furthermore it would highly depend on this preferred 
coordinate system.

For the above problem \emph{Philip E. Parker} suggested us a partial solution, 
which avoids coordinate systems. 
He drew our attention to his work [1], which partly deals with a problem of fall-off properties. 
In his paper, he uses the topological approach to infinities of manifolds: 
the set of \emph{ends} of the manifold $C$ can be defined as $E(C):=\underset{\underset{\text{compact}}{K\subset C}}{\mathrm{liminv}}\;\pi_{0}(C\setminus K)$, where 
$\pi_{0}(C\setminus K)$ means the set of the connected components of $C\setminus K$, and 
$\mathrm{liminv}$ is the so called inverse limes, known in topology. 
An end represents an infinity in the topological sense. Then, he is able to 
define when two Riemannian metric tensor fields (of some vector bundle) falls off at a given 
infinity in the same way (notion of order relatedness). This notion provides 
an equivalence class concept between Riemannian metric tensor fields. Given such an equivalence class of Riemannian metric tensor fields, 
one can define the notion of \emph{rapidly decreasing} field configurations, 
which can be used to introduce fall-off properties. 
However, as indicated, this concept highly depends on the used metric tensor field equivalence class, the 
(physical) meaning of which is quite unclear (just as in the case of preferring a global 
coordinate system on $C$). Furthermore, the method would also highly depend on the 
initial splitting of the spacetime manifold into $\mathbb{R}\times C$, which 
conflicts with the philosophy of the theory of relativity.

\section{Discussion}

We have seen, that the variational formulation of general relativistic field theories 
can be defined with a significant mathematical elegance over compact 
base manifolds (with or without boundary). Over noncompact base manifolds, 
the variational principles can be defined with a great effort, the known 
constructions are not elegant at all in mathematical sense, furthermore they 
do have problems with the interpretation.\footnote{The problem 
of \emph{non-constructiveness} 
in the case of one-parameter family constructions, furthermore the problem of \emph{spacetime 
splitting} and the \emph{metric tensor field equivalence class dependence of fall-off properties} 
in the case of time-slice constructions.}

In physics, it is held as a principle, that the equation of motion of fields 
arise from some Euler-Lagrange equations (that is, as some equation $DS(v,\nabla)=0$). If 
we want to preserve this principle, and want to avoid the rather questionable 
constructions in the noncompact cases at the same time, we can make a choice 
to solve the problem.

\begin{enumerate}
\item We can restrict the spacetime models to compact orientable cases.
\item We do not interpret the base manifold as the spacetime manifold itself, 
but as a kind of compactification of it.
\end{enumerate}

The first case is unacceptable: it is a theorem, that every compact spacetime 
model admits closed timelike curves. So, a compact spacetime model, arising from 
any kind of formulation, cannot be considered physically realistic.

The other case does not have physical obstructions, and has a certain mathematical elegance. 
But then, the question arises: if we do not interpret the base manifold directly 
as the spacetime manifold, how do we interpret it?

For this problem, a possible solution is the condition of asymptotic simpleness 
of a spacetime. (See e.g. [5], [8].) If this condition holds, then one can define 
the notion of conformal infinities of the spacetime and the conformal compactification 
of the spacetime, which will be a compact manifold with boundary.

From the above argument, it is likely to consider only the compact case of a 
base manifold (with boundary), and interpret it as the conformal compactification 
of the spacetime manifold.

\begin{Thm}
Let the base manifold $M$ be compact with boundary. Let 
\[(v,\nabla)\in\Gamma^3(\breve{V}(M))\times\breve{D}^3(T(M),V(M)).\] 
Then the condition $DS(v,\nabla)=0$ is equivalent to the followings: 
\begin{enumerate}
\item the \emph{Euler-Lagrange equations}, that is the equations 
\[D_{1}\mathbf{dL}(v,\nabla v,F_{\nabla})-(\nabla{\cdot}D_{2}\mathbf{dL}(v,\nabla v,F_{\nabla}))-(\Tr{T_{\nabla}}{\cdot}D_{2}\mathbf{dL}(v,\nabla v,F_{\nabla}))=0,\]
\[D_{2}\mathbf{dL}(v,\nabla v,F_{\nabla})(\cdot)v-2(\nabla{\cdot}\widehat{D_{3}\mathbf{dL}}(v,\nabla v,F_{\nabla}))(\cdot)-2(\Tr{T_{\nabla}}{\cdot}\widehat{D_{3}\mathbf{dL}}(v,\nabla v,F_{\nabla}))(\cdot)=0\]
are satisfied on the interior of $M$, 
and
\item the \emph{boundary constraints}, that is the equations 
\[\Tr(D_{2}(\mathbf{dL}(v,\nabla v,F_{\nabla}))=0,\]
\[\Tr(2\widehat{D_{3}\mathbf{dL}}(v,\nabla v,F_{\nabla}))(\cdot)=0\]
are satisfied on the boundary of $M$.
\end{enumerate}
\end{Thm}

\begin{Prf}
Let us take such sections $({\delta}v,{\delta}C)$, that their support is in the interior of $M$. 
Then, the boundary term is zero in the Euler-Lagrange relation in \emph{Remark 14}. 
By the Lagrange lemma, condition \emph{1} is implied.

We know now, that condition \emph{1} holds. This means that the non-boundary term in the 
Euler-Lagrange relation in \emph{Remark 14} is zero. Now taking any sections $({\delta}v,{\delta}C)$, 
condition \emph{2} is implied, by using the Lagrange lemma on the boundary of $M$.
\end{Prf}

The question arises: what do the boundary conditions and the boundary of the base manifold 
mean? In the next section, we shall investigate the physical meaning of the boundary 
conditions on the example of empty general relativistic spacetime: we shall 
show that the boundary represents the conformal boundary (conformal infinity) of the arising 
spacetime model.

\section{Boundary as conformal infinity: the example of empty general relativistic spacetime}

Let the base manifold $M$ be $4$ dimensional, and let us require that $M$ 
admits $C^3$ semi-Riemann metric tensor fields with Lorentz signature 
(this is known to hold 
if and only if there exists a nowhere zero $C^3$ tangent vector field on $M$). 

Let us take the vector bundle $V(M):=F(M){\times}\overset{2}{\vee}T^{\ast}(M)$ ($\vee$ means 
symmetrized tensor product). 
We define the sub fiber bundle $\breve{V}(M)$ by the restriction of the fibers 
of $V(M)$ in the following way: for each point $p\in M$ the fiber is restricted 
to $\mathbb{R}{\times}L_{p}(M)$, where $L_{p}(M)$ denotes the subset of 
semi-Riemannian metric tensors with Lorentz signature in $\overset{2}{\vee}T^{\ast}_{p}(M)$. 
It can be easily shown, that $\breve{V}(M)$ is such a sub fiber bundle of the vector 
bundle $V(M)$, as required in the text.

Let us take the sub affine space of $D^{3}(T(M)){\times}D^{3}(V(M))$, which has 
the following property: the sub affine space should consist of those pairs $(\nabla,\nabla^{'})$, where 
the covariant derivation $\nabla^{'}$ over $V(M)$ corresponds to the covariant derivation 
obtained by the unique extension of $\nabla$ 
to $F(M){\times}\overset{2}{\vee}T^{\ast}(M)$, by using \emph{Remark 1}. 
This sub affine space can be naturally identified with $D^{3}(T(M))$, therefore we can 
define a covariant derivation from this sub affine space to be torsion-free if and only 
if the corresponding covariant derivation from $D^{3}(T(M))$ is torsion-free. 
Let $\breve{D}^{3}(T(M),V(M))$ 
be the sub affine space of torsion-free covariant derivations of the previous sub 
affine space. It can be easily shown, that this is a closed sub affine space with respect 
to the topology defined in \emph{Definition 8}.

In this subsection we will apply the usual formalism of Penrose 
abstract indices, to denote tensor quantities and various contractions of them.

If $g_{ab}(p)$ is a metric tensor with Lorentz signature from $\overset{2}{\vee}T^{\ast}_{p}(M)$ ($p{\in}M$), 
then the inverse metric of it (the corresponding Lorentz metric in $\overset{2}{\vee}T_{p}(M)$) 
will be denoted by $g^{ab}(p)$. Let us take an orientation of $M$. 
One of the two associated volume forms to a $g_{ab}(p)$ Lorentz metric (that one, which corresponds to the chosen orientation) 
will be denoted by $\mathbf{dv}_{g}(p)$.

If $\nabla$ is a covariant derivation on $T(M)$, the corresponding Riemann-tensor 
will be denoted by $R_{\nabla}$.

With the above notations, let us take the Lagrange form 
\[\mathbf{dL}: ((\varphi,g_{ab}),(D\varphi,D{g_{cd}}),({R_{efg}}^{h}))\mapsto\mathbf{dv}_{g}\varphi^{2}g^{ik}{\delta}^{j}_{l}{R_{ijk}}^{l},\]
which is the abstraction of the Einstein-Hilbert Lagrangian. 
The field $\varphi$ will play the role of the geometrized coupling factor to 
gravity, that is the inverse of the Planck length.

\begin{Thm}
The Euler-Lagrange equations of the present field theory are 
\[2\mathbf{dv}_{g}{\varphi}g^{ac}{\delta}^{b}_{d}{(R_{\nabla})_{abc}}^{d}=0,\]
\[-\mathbf{dv}_{g}\varphi^{2}\Bigl(g^{ae}g^{fc}{\delta}^{b}_{d}{(R_{\nabla})_{abc}}^{d}-\frac{1}{2}g^{ef}g^{ac}{\delta}^{b}_{d}{(R_{\nabla})_{abc}}^{d}\Bigr)=0,\]
\[-\nabla_{a}\Bigl(\mathbf{dv}_{g}\varphi^{2}(g^{ac}{\delta}^{b}_{d}-g^{bc}{\delta}^{a}_{d})\Bigr)-(T_{\nabla})_{ae}^{e}\Bigl(\mathbf{dv}_{g}\varphi^{2}(g^{ac}{\delta}^{b}_{d}-g^{bc}{\delta}^{a}_{d})\Bigr)=0,\]
which hold in the interior of $M$.

The boundary constraints are 
\[0=0,\]
\[0=0,\]
\[(\mathbf{dv}_{g})_{afgh}\varphi^{2}(g^{ac}{\delta}^{b}_{d}-g^{bc}{\delta}^{a}_{d})=0,\]
which hold on the boundary of $M$.
\end{Thm}

\begin{Prf}
One can get these equations, by simply substituting $\mathbf{dL}$ into the formulae in 
\emph{Theorem 15}, and by using the identities $\frac{\partial{\mathbf{dv}_{g}}}{\partial{g_{ef}}}=\frac{1}{2}g^{ef}\mathbf{dv}_{g}$ and $\frac{\partial{g^{ac}}}{\partial{g_{ef}}}=-\frac{1}{2}(g^{ae}g^{fc}+g^{af}g^{ec})$, 
which can be derived easily, but also can be found in [5] or [8].
\end{Prf}

It is easily seen, that the first Euler-Lagrange equation follows from the second one on the domains, where $\varphi$ is nowhere zero. 
Furthermore, the first two boundary constraint is trivial. 
Let us denote the torsion-free part of $\nabla$ with $\tilde{\nabla}$. Then the 
third Euler-Lagrange equation is equivalent to the equation $-\tilde{\nabla}_{a}\Bigl(\mathbf{dv}_{g}\varphi^{2}(g^{ac}{\delta}^{b}_{d}-g^{bc}{\delta}^{a}_{d})\Bigr)=0$.

\begin{Lem}
On those open sets, where $\varphi$ is nowhere zero, the equation 
\[-\tilde{\nabla}_{a}\Bigl(\mathbf{dv}_{g}\varphi^{2}(g^{ac}{\delta}^{b}_{d}-g^{bc}{\delta}^{a}_{d})\Bigr)=0\]
is equivalent to the equation $\tilde{\nabla}_{a}(\varphi^{-2}g^{bc})=0$.
\end{Lem}

\begin{Prf}
The proof will be performed separately in the two implication directions.

$(\Leftarrow)$ This way is trivial, it can be shown by direct substitution.

$(\Rightarrow)$ To prove this way, let us contract the first equation in its $b$ and $d$ indices. 
We get $-\tilde{\nabla}_{a}\Bigl(\mathbf{dv}_{g}\varphi^{2}3g^{ac}\Bigr)=0$. Therefore, 
the first equation implies $\tilde{\nabla}_{a}\Bigl(\mathbf{dv}_{g}\varphi^{2}g^{bc}\Bigr)=0$. 
Let us introduce the rescaled metrics $G_{ab}:=\varphi^{2}g_{ab}$ and $G^{ab}:=\varphi^{-2}g^{ab}$. 
Then the implied equation can be written into the form $\tilde{\nabla}_{a}\Bigl(\mathbf{dv}_{G}G^{bc}\Bigr)=0$. 
It can be easily seen, for example by using coordinates and the relation $\frac{\partial{\mathbf{dv}_{g}}}{\partial{g_{ef}}}=\frac{1}{2}g^{ef}\mathbf{dv}_{g}$, 
that $\tilde{\nabla}_{a}\mathbf{dv}_{G}=\frac{1}{2}\mathbf{dv}_{G}G^{bc}\tilde{\nabla}_{a}G_{bc}$ for 
arbitrary covariant derivatives $\tilde{\nabla}$, from which, by the Leibniz rule we infer that 
$\mathbf{dv}_{G}(\tilde{\nabla}_{a}G^{bc}-\frac{1}{2}G^{bc}G_{de}\tilde{\nabla}_{a}G^{de})=0$. 
We can drop $\mathbf{dv}_{G}$ from this equation, because it is nowhere zero on the 
domain in question. Furthermore, by taking its contraction with $G_{bc}$, 
one gets $-G_{de}\tilde{\nabla}_{a}G^{de}=0$. Therefore, by using this and the previous 
equation: $\tilde{\nabla}_{a}G^{bc}=0$, so finally $\tilde{\nabla}_{a}(\varphi^{-2}g^{bc})=0$ 
is implied.
\end{Prf}

We can summarize now the Euler-Lagrange equations: they are equivalent to 
\[\tilde{\nabla}_{a}(\varphi^{2}g_{bc})=0,\]
\[\varphi^{2}\Bigl({(R_{\nabla})_{acb}}^{c}-\frac{1}{2}(\varphi^{2}g_{ab})(\varphi^{-2}g^{ef}){(R_{\nabla})_{ecf}}^{c}\Bigr)=0\]
on those domains of the interior of $M$, where $\varphi$ is nowhere zero. 
From the definition of $\breve{V}(M)$ and $\breve{D}^{3}(T(M),V(M))$ we know, that 
$\varphi^{2}g_{ab}$ has Lorentz signature on the above domains, furthermore $\tilde{\nabla}=\nabla$. 
Thus, the \emph{vacuum Einstein equations} turned out to be equivalent to the 
Euler-Lagrange equations on those domains of the interior of $M$, where the field 
$\varphi$ is nowhere zero.

We can summarize the boundary constraints: they are equivalent to 
$\varphi|_{\partial{M}}=0$, which is also equivalent to $(\varphi^{2}g_{ab})|_{\partial{M}}=0$. 
The latter means that the boundary of $M$ 
is the conformal infinity of the arising spacetime model.

The rescaled metric $\varphi^{2}g_{ab}$ can be interpreted physically, as the metric, measured 
in such units, where the coupling factor of gravity (that is, the inverse of the Planck length) 
is taken to be $1$.

\section{Concluding remarks}

A mathematically precise global approach was presented, to obtain variational 
formulation of general relativistic classical field 
theories. According to the authors' information, 
there is no such formulation, known in literature. For an overview of variational 
principles over general relativity, see [6]. For further recent works in 
topic, see [2], [3], [4], [7].

The development of such a formulation was inspired by some problems of the 
usual approaches, and by possible future application as a tool 
in the proof of a global existence theorem: as the approach is global, one would 
simply have to prove critical point theorems on the action functional in order to 
obtain a theorem on the global existence of solutions. This would be desirable, as 
the question of global existence of solutions is unsolved in general 
relativistic field theories, yet.

\section*{Acknowledgements}

I would like to thank Tamás Matolcsi and János Szenthe, for discussions, 
for reading through the versions of this article, and for suggesting several 
corrections.

I would like to express my special thanks to Tamás Matolcsi, for providing me 
deeper insights to physics, by using mathematics.

Furthermore, I would like to than to Philip E. Parker for his valuable 
remark concerning the treatment of the fall-off properties of field configurations.

\section*{References}

\def\inum#1{[#1]}

\begin{description}
\item[\inum{1}] John K. Beem, Philip E. Parker: \emph{The Geometry of Bicharacteristics and Stability of Solvability}; in Differential Geometry, Calculus of Variations and Their Applications, edited by George M. Rassias and Themistocles M. Rassias, 1985, 5, p. 83-94
\item[\inum{2}] R. D. Bock: \emph{Local Scale Invariance and General Relativity}; Int. J. Theor. Phys., 2003, 42, p. 1835-1847
\item[\inum{3}] J. D. Brown, J. W. Jork: \emph{Action and Energy of the Gravitational Field}; To be published in Annals of Physics
\item[\inum{4}] F. Embacher: \emph{Actions for Signature Change}; Phys. Rev., 1995, D51, p. 6764-6777
\item[\inum{5}] S. W. Hawking, G. F. R. Ellis: \emph{The Large Scale Structure of Space-Time}; Cambridge University Press, 1973
\item[\inum{6}] P. Peldán: \emph{Actions for Gravity, with Generalizations: a Review}; Class. Quant. Grav., 1994, 11, p. 1087-1132
\item[\inum{7}] J. M. Pons: \emph{Boundary Conditions from Boundary terms, Noether Charges and the Trace K Lagrangian in General Relativity}; Gen. Rel. Grav., 2003, 35, p. 147-174
\item[\inum{8}] R. M. Wald: \emph{General Relativity}; University of Chicago Press, 1984
\end{description}

\end{document}